\documentclass[12pt]{iopart}
\usepackage{iopams}
\jl{6}

\def\st{\rm{ st}}
\def\R{\mathbb{R}}                

\def\cross{\times}                
\def\id{\mathrm{id}}	          
\newtheorem{Proposition}{Proposition}[section]

\newtheorem{definition}[Proposition]{Definition}
\newtheorem{remark}[Proposition]{Remark}
\def\M{\ensuremath{\mathcal{M}}}   
\def\RO{\rho}                      
\def\Rot{R}                        
\def\PR{\mathcal{R}}               

\begin{document}
\title{Spin half in classical general relativity} 
\author{Mark J Hadley}

\address{Department of Physics, University of Warwick, Coventry
CV4~7AL, UK\\ email: Mark.Hadley@warwick.ac.uk}

\begin{abstract}

It is shown that models of elementary particles in classical general
relativity (geons) will naturally have the transformation properties
of a spinor if the spacetime manifold is not time orientable. From a
purely pragmatic interpretation of quantum theory this explains why
spinor fields are needed to represent particles. The models are based
entirely on classical general relativity and are motivated by the
suggestion that the lack of a time-orientation could be the origin of
quantum phenomena.

\end{abstract}

\maketitle

\section{Introduction} \label{introduction}

Spin half is normally regarded as a non-classical property peculiar to
quantum theory. Objects with spin-half are represented by spinor
fields. The most distinctive, indeed the characteristic, property of a
spinor is the way it transforms under a rotation. In particular the
fact that a rotation of space coordinates by an angle of $2 \pi$ is
not an identity operation whereas a rotation by $4 \pi$ is an identity
operation. This work offers an explanation of why spinor fields are
needed to represent fermions. The explanation is based entirely upon
classical general relativity.

Many authors have explored the relation between general relativity and
quantum theory, some attempt to apply quantum theory to curved
spacetime and non-trivial spacetime topologies by adding quantum
fields to the underlying manifold. Such an approach with spin-half
particles requires the manifold to be endowed with a spinor field;
Geroch \cite{geroch68} describes the constraints on the topology and
orientability of spacetime in order to admit a spinor field. This work
does not attempt to add or construct a spinor field on a curved
spacetime. Instead it attempts to explain why we need to use spinor
fields (on flat spacetime) to model some particles. The constraints
imposed by Geroch do not apply.

In quantum theory, a wavefunction, or quantum field, is used to
calculate the probabilistic results of experiments which detect a
particle. This is certainly true. To go beyond this pragmatic view
requires an interpretation of quantum theory (see for example
Ballentine \cite{ballentine} or Isham \cite{isham95} for a
discussion). To regard the wavefunction itself as being the particle
has substantial conceptual and practical problems. But if the particle
and wavefunction are indeed different, then a model or description of
an elementary particle is required. One way of modelling an elementary
particle is as a region of space-time with non-trivial topology - a
geon \cite{einstein_rosen,misner_wheeler} which can be described using
classical general relativity. The wavefunction would still be the tool
for calculating the results of experiments but its role would be that
of a probability function.  It is important to keep in mind this
distinction between the particle itself and the quantum fields used to
calculate the results of experiments.

The complex fields used to represent particles need to have, amongst
other attributes, the same transformation properties as the particles
that they represent. Particles with zero spin and spherical symmetry
are represented by a complex scalar field. Spin-1 particles with the
transformation properties of a vector are represented by vector
fields. Note that vectors and spinors are defined by their
transformation properties under rotations. An elementary particle (or
indeed any object) may have the transformation properties of a vector
- a pencil is a good approximation. The vector would describe its
orientation, and a rotation of either the object or the coordinate
system would be correctly modelled by the vector. By contrast the
quantum mechanical wavefunction defines a vector (or scalar) at each
point in space (a vector field) and the components would be complex
numbers - this is what quantum theory requires to compute the
probabilistic results of experiments. In a similar way there is a
distinction between a single object having the transformation
properties of a spinor and a spinor field which is used by quantum
mechanics to represent such an object.

Some authors have produced artificial models, parts of which,
transform as a spinor see for example the cube within a cube described
by Misner, Thorne and Wheeler\cite{MTW} or the tethered rocks of
Hartung \cite{hartung}. This paper shows how an object with the
transformation properties of a spinor can be constructed within
classical general relativity. The work requires that the topology of
spacetime within a fermion is non-trivial and is not time
orientable. It would then follow that quantum theory in flat spacetime
would need to model the dynamics of such a particle using spinor
fields.

This paper relates to models of elementary particles as regions of
spacetime with non-trivial topology. The term geon was used by Misner
and Wheeler to describe particle-like structures with non-trivial
spatial topology \cite{misner_wheeler}. The idea was later extended by
Hadley to regions which also had non-trivial causal structure called
4-geons \cite{hadley97}. The investigation of manifolds which are not
time orientable is motivated by the suggestion that models of
elementary particles based on such manifolds would give rise to
quantum mechanical effects \cite{hadley97} and that a reversal of the
time orientation is related both to topology change \cite{hadley98}
and the measurement process in quantum theory. Therefore, the
manifolds being examined have real significance quite apart from their
transformation under rotations. The simplest example was described in
the paper by Diemer and Hadley \cite{diemer_hadley} and shown to
display net charge from the source-free Maxwell equations. Hadley's
gravitational explanation for quantum theory provides additional
motivation for this work, but the results of this paper are
interesting in their own right and do not depend upon the earlier
work. The main result is a fascinating consequence of non-trivial
causal structure in general relativity.

Here it is shown that non time orientable manifolds present
obstructions which prevent a physical (time parameterised) rotation
being extended throughout the manifold in a non-trivial way. There
must be some exempt points. This topological obstruction will, in
general, prevent a $2 \pi$ rotation (and any well behaved extension of
it) from being an identity operation. This property of spinors is
intimately related to the fact that the group of rotations, SO(3), is
not simply connected. A result which can be seen in the famous
scissors model of Dirac and the cube within a cube described by
Misner, Thorne and Wheeler \cite{MTW}. A description is given in
references \cite{feynman_weinberg, penrose_rindler} and \cite{newman}
contains a proof.

Friedman and Sorkin \cite{friedman_sorkin} have previously considered
other topological obstructions which prevent a rotational
transformation being extended from the region of flat spacetime
throughout the manifold (ie into the interior of the particle). On
some 3-manifolds (characterised by Hendricks \cite{hendriks}) it is
not possible to define a rotational vector field. In the context of
quantum gravity, Friedman and Sorkin impose a wavefunction on such
manifolds and consider the effects of a rotation on the asymptotically
flat regions. They conclude that a wavefunction as required by a
theory of quantum gravity could have a spinorial character. By
contrast; this work invokes a different mechanism, it applies to
manifolds which have a physical interest for other
reasons. Furthermore, these results arise from classical general
relativity, they do not invoke any quantum theory of gravity, but they
may help to explain quantum phenomena.

\section{Line bundle models of spacetime}
Although the analysis requires a clear distinction between time and
space directions, the results can be obtained by applying topological
arguments to a non-trivial line bundle over a 3-manifold - it is
not necessary to define a metric explicitly.

A causal spacetime is modelled by a trivial line-bundle $E$ over a
3-manifold, \M, $ E \stackrel{\pi}{\rightarrow} \M$. The base
manifold, \M, corresponds to space at $t = 0$. The evolution is given
by a diffeomorphism, $Q$ from $\M \cross \R$ to the line bundle $E$
such that $Q(\cdot ,0)$ is the zero section and $(Q(\cdot , t))(\M)$
is the space at a time $t$. Under a transformation a fixed point in
space would be mapped to a point on its own fibre. A non-trivial
line-bundle over a 3-manifold is a convenient topological model of
those non-time orientable spacetimes with the property that time is
not orientable on any spacelike slice; the manifold described by
Diemer and Hadley is an example.

A base 3-manifold endowed with a non-trivial line bundle would
correspond to a spacetime that is not time orientable. If the line
bundle is non-trivial then every section has some zero points. In
general, the space of zero points, $X$, is a 2-dimensional surface. It
is this topological property of sections of a non-trivial line bundle
which gives rise to the results in this paper. This can be illustrated
by considering sections of a m\"obius strip. The base manifold is a
circle, $S^1$, which corresponds to a circular space - the time
direction is everywhere normal to the base circle. Parts of the base
manifold can be moved within the M\"obius band, but an attempt to
shift the entire circle up or down is not possible - there will be at
least one point where the shifted circle crosses the base circle.

\section{Models of particles}

A particle in spacetime is modelled as an asymptotically flat
3-manifold with non-trivial topology and endowed with a line bundle
where each fibre corresponds to the timeline of a point in space. A
non-time-orientable manifold corresponds to a non-trivial line
bundle. 

The simplest example due to Diemer and Hadley\cite{diemer_hadley} is
RP3 with a point removed and endowed with a non-trivial
line-bundle. The point removed corresponds to spatial infinity and a
sphere $S^2$ enclosing the point corresponds to a sphere enclosing the
particle. With the addition of a Faraday electromagnetic field 2-form,
this construction can model a spherically symmetric electric monopole.

An alternative construction of the same spacetime manifold is as
follows: a ball is removed from space (a cylinder from spacetime) and
then joined up by identifying opposite points, but reversing the time
direction. Externally this is spherically symmetric. The region of
non-trivial topology can be surrounded by a sphere, $S^2$, such that
spacetime is topologically trivial and asymptotically flat outside the
sphere.

\section{Definitions of rotations} 
Rotations are well-defined in $\R^3$ as elements of SO(3). Every
rotation has an axis, $\zeta$, and is an element of the one parameter
subgroup of rotations about this fixed axis. Rotations,
$\RO_\zeta(\theta)$, by an angle $\theta$ about an axis $\zeta$ are
smooth maps from $\R^3$ to $\R^3$ which satisfy the following
conditions:

\begin{definition}[1-parameter subgroup conditions]
\label{def:cont}\mbox{}
\begin{enumerate}
\item $\RO_\zeta(0) = \id$ 
\item $\forall \theta,\phi \in \R\ \RO_\zeta(\theta + \phi) =
\RO_\zeta(\theta)\RO_\zeta(\phi)$
\item $\RO_\zeta$ is a smooth function
\end{enumerate}
\end{definition}

In the following the subscript $\zeta$ will be omitted for clarity.

We need to extend the usual definition so that it applies to manifolds
with a non-trivial topology and to manifolds which are not
time-orientable. For physical reasons we restrict attention to
asymptotically flat manifolds, and in the asymptotically flat region
we require correspondence with rotations defined in $\R^3$.

\begin{definition}[A Rotation on an Asymptotically Flat Manifold]
\label{def:asmrotation}A rotation, $\Rot(\theta)$, by an angle
$\theta$, on an Asymptotically Flat Manifold, \M, is an element of a
1-parameter subgroup of diffeomorphisms from \M\ to itself which
satisfies the conditions \ref{def:cont} above and $\forall \theta \in
\R$ converges to a rotation, $\RO_\zeta(\theta)$, in $\R^3$ for large
$x$:
\begin{equation}
{\rm dist}(\RO_\zeta(\theta)x, \Rot(\theta)x) \rightarrow 0
\hspace{10mm}\hbox{\rm as \hspace{5mm}$|x|\rightarrow \infty $ }
\end{equation}
\end{definition}

The definition is very wide, the form of the transformation is only
tightly prescribed in the asymptotic regions. In particular
$\Rot(\theta)$ is not an isometry, it is any smooth extension of
the familiar form of a rotation.
\begin{remark} 
A rotation always exists - consider two concentric spheres $S^2_a$ and
$S^2_b$ where $S^2_b$ encloses $S^2_a$ which in turn encloses the
region of non-trivial topology:
\begin{equation}
\Rot(\theta) = \left\{ \begin{array}{ll}
                     \id = \Rot(0)& \hbox{\rm inside $S^2_a$}\\
                     \RO(\theta) & \hbox{\rm outside $S^2_b$}\\
		     \RO(f(r)\theta)&\hbox{\rm in between} 
                  \end{array}
               \right.
\end{equation}
where $f(r)$ is any smooth function of $r$ such that $f(r)= 0$ when $r=
r_a$ and $f(r) = 1$ when $r= r_b$.
\end{remark}

\begin{remark}
Rotations are not uniquely defined by
definition~\ref{def:asmrotation}.  Any smooth function $f(r)$ with the
properties above can be used to construct a rotation $\Rot(\theta)$
which satisfies \ref{def:asmrotation} and has the same asymptotic
form.
\end{remark}

A rotation is an element of a 1-parameter subgroup of diffeomorphisms
acting on \M. The rotation and the subgroup define a path in \M\ from
$x$ to $\Rot(\theta)x$: $\gamma_\theta(\lambda) = \{\Rot(\lambda
\theta)x| x \in \M, \lambda \in[0,1]\}$, This definition has no notion
of time. The parameter, $\lambda$ has no physical significance, the
rotation is not an operation that can be physically implemented. Of
specific concern in this paper is the notion of a physical rotation -
a rotation which defines a worldline parameterised by time. This gives
an operation that is physically relevant, an object or a 3-manifold
can be transformed from an initial state at $t=0$ to a rotated state
at a later time $t=1$:

\begin{definition}[Physical Rotation on a time-orientable manifold]
\label{def:PR_orient} A physical rotation, $\PR(\theta)$ on a
time-orientable manifold is a map from \M, identified with the zero
section, to the image of the unit section of the line bundle:
\begin{equation}
\PR(\theta)(x,0) \rightarrow (\Rot(\theta)x,1)
\end{equation}
\end{definition}
where $\PR(\theta)$ also satisfies the 1-parameter subgroup conditions
\ref{def:cont}. The projection of a physical rotation onto the base
manifold gives a rotation of an asymptotically flat manifold defined
earlier:
\begin{equation}
\label{eq:project}
\pi \circ \PR(\theta) = \Rot(\theta)
\end{equation}

A physical rotation and its subgroup define a curve in the line bundle
$E$ from $(x,0)$ to $(\Rot(\theta)x,1)$,
$\chi_\theta(t)=\{(\Rot(t\theta)x,t)| x \in \M, t\in[0,1]\}$, which is
the worldline of each point $x \in \M$. The projection of the
worldline is equal to the path in \M\ defined earlier: $\pi \circ
\chi_\theta(t) = \gamma_\theta(t)$. Although a physical rotation is
conceptually different, there is a one to one correspondence between
physical rotations and rotations of an asymptotically flat manifold
defined earlier. Similarly the curves, $\gamma_\theta(\lambda)$, and
world lines, $\chi_\theta(t)$ are in one to one correspondence.

However definition \ref{def:PR_orient} cannot be applied to a
spacetime which is not time orientable (a non-trivial line bundle)
because it assumes the existence of a nowhere vanishing section (the
unit section) which implies that a global trivialisation of the line
bundle exists. We refine definition~\ref{def:PR_orient} so that it
applies to any asymptotically flat manifold including those which are
not time orientable. The definition requires the replacement on the
unit section of the line bundle by another section $\phi$ :

\begin{definition}[Physical Rotation]
\label{def:physicalR}
A physical rotation, $\PR(\theta)$ is a map from \M, identified with
the zero section, to the image of a section $\phi$ of the line bundle:

\begin{eqnarray}
\PR(\theta)(x,0) &\rightarrow &(\Rot(\theta)x,\phi(x))\\ 
\mbox{\rm when } \phi(x) = 0&&\Rot(\theta)x = x\ \forall \theta
\label{eq:exempt_point}\\ 
\mbox{\rm as } |x|\rightarrow \infty &&\phi(x) \rightarrow 1
\end{eqnarray}
\end{definition}

Clearly a physical rotation exists and is not
unique. Equation~\ref{eq:project} is still valid but there is no
longer a one to one correspondence between physical rotations and
rotations of an asymptotically flat manifold due to
equation~\ref{eq:exempt_point}. 

This definition is obviously of physical relevance. A real rotation in
the laboratory is well-defined in an almost flat region of
spacetime. The internal structure of a particle is unknown and the
structure of spacetime within a particle is also unknown. This
definition places few constraints on the internal structure of a
particle and accommodates both a spacetime with non-trivial topology
as well as the existence of extraneous fields on a flat spacetime.

In general a {\em Physical Rotation} is not an isometry, it does not
preserve distances except in the asymptotic region. It is defined as a
general extension of a rotation from the asymptotic region to the
whole manifold. An isometric extension would not normally be
possible. An analogy would be a large rubber disk, fixed in the middle
with two or more nails. The outer rim could be rotated, near the outer
rim the rotation would be an isometry. The transformation created by
rotating the rim, can be extended to the whole disk, but it is not
an isometry throughout the disk.

This construction agrees with the normal definition of a rotation in
the asymptotically flat region, it can be extended to the whole of a
time orientable manifold with $\phi(x) =1 \forall x$. It can also be
extended to the whole manifold even when it is not time orientable,
but for some points $\phi(x) =0$. Definition~\ref{def:physicalR} and
equation~\ref{eq:exempt_point} distinguish two types of fixed points
under a rotation.

\begin{definition}[fixed point]
$x$ is a fixed point of the rotation if $\PR(\theta) (x,0) \rightarrow
(x,\phi(x))$
\end{definition}
These points correspond to the axes of a rotation. But there are also
exempt points:
\begin{definition}[exempt point]
$x$ is an exempt point of the rotation if $\PR(\theta) (x,0)
\rightarrow (x,0)$
\end{definition}
We denote by $X$ the set of all exempt points. The name exempt
conforms to the definition of Hartung\cite{hartung}, where he
considers the rotation of a tethered object, the object rotates but
the other end of the tether is an exempt point in the sense of being a
point to which the rotation is not applied.

As already pointed out, rotations are not uniquely specified by this
definition. The rotation is tightly defined in the asymptotic region
where it maps $(x,0) \rightarrow (\Rot(\theta)x,\phi(x)) \approx
(\RO(x),1)$, but there are many ways of extending the section of the
line bundle and many ways of extending the rotation
transformation. Consequently the exempt points are not determined by
the rotation in the asymptotic region, but depend upon the way in
which the section is extended. On a manifold which is not
time-orientable, time is not orientable around at least one class of
non-contractable closed curves in \M \ and therefore every such curve
must have at least one exempt point. Consequently, the space of exempt
points is at least two dimensional (otherwise a small distortion of
the curve could be made which would not have an exempt point) and the
proof in \cite{newman} applies.

\section{Spin-half}
The following results apply to the specific particle model described
above and also to the more general case where the particle is a region
of spacetime with non-trivial topology surrounded by a world tube $S^2
\cross \R$ such that any region spanning the $S^2$ does not admit a
time orientation. This is sufficient to ensure that any section of the
line bundle has exempt points.

The fact that the exempt points form a 2-dimensional surface ensures
that this model is analogous to the tethered rocks of Hartung
\cite{hartung} and the scissors trick of Dirac
\cite{feynman_weinberg}. It follows that a $2 \pi$ rotation cannot be
an identity operation but $4 \pi$ rotation can be:
\begin{eqnarray}
\forall \PR(0)  &(x,0) \rightarrow (x,\phi(x))& \\
\forall \PR(2 \pi)\ \exists x\  \st: &(x,0) \rightarrow
 (\Rot(2 \pi)x,\phi(x))& \ne (x,\phi(x))\\
\exists \PR(4 \pi)\ \st:\ \forall x\ &(x,0) \rightarrow 
(\Rot(4 \pi)x,\phi(x))& = (x,\phi(x))
\end{eqnarray}

It follows that $\PR(0)$, the identity transformation, is
diffeomorphic to $\PR(4\pi)$ relative to $\{X,S^2\}$ and that $\PR(2
\pi)$ is not. This is also a necessary, but not sufficient condition,
for $\PR(4\pi)$ to be an isometry and for $\PR(2\pi)$ not to be an
isometry. A $2\pi$ rotation could be an isometry if the particle had
an internal symmetry - which is not possessed by the simple monopole
example.

This gives a physical explanation of why it is appropriate to model a
fermion as a tethered object. This model, with non-trivial causal
structure, was originally constructed to explain how the effects of
quantum theory could arise within classical general relativity. The
spin half effects (and also the appearance of electric charge
\cite{diemer_hadley}) are simply a consequence of using general
relativity to explain quantum effects.

\section{Conclusion}
On the relation to quantum theory, we take a pragmatic view of quantum
theory {\em that it is a scheme for predicting the probabilistic
distribution of the outcomes of measurements}.  It is clear that a
particle must be represented by a mathematical object with the
appropriate transformation properties under rotations. The particle
modelled here has the transformation properties of a spinor, so any
field theory which attempts to model the behaviour under rotations
will necessarily use spinor fields.

This approach contrasts with that of Sorkin, who considered a
wavefunction imposed upon a non-trivial manifold. That would be
meaningless when the constructions given can display quantum
mechanical effects by themselves. Here the wavefunction is defined in
$\R^4$ and is a means of mapping the evolution of non-trivial causal
structure onto a conventional spacetime.

The previous sections modelled a single free particle. Hartung extends
the idea of a tethered rock to two or more tethered rocks and
concludes that an exchange of particles is equivalent to a $2 \pi$
rotation of one of the particles. However his analysis assumes that
two tethered rocks is equivalent to a pair of rocks tethered to each
other. With the model presented here, such an equivalence is not
apparent.

In general relativity, the {\em non-gravitational} energy, momentum
and angular momentum can be derived from the energy momentum tensor
which is defined at every point of spacetime. However the contribution
of the gravitational field itself to the total energy, momentum and
angular momentum cannot be defined locally and hence the total values
cannot be defined locally (see for example~\cite{MTW}). Well defined
global values exist in asymptotically flat manifolds such as the one
described here. This example has a total angular momentum of zero. The
asymptotic form of the metric is independent of the time
orientability. This must be the case because a non-zero value would
give a well-defined spin direction independently of measurement which
is contrary both to quantum theory and experiment. Indeed a zero
angular momentum for an unpolarised electron is just what one would
expect from the Bohm interpretation of quantum mechanics.

In conclusion, the transformation properties of fermions have been
modelled using classical general relativity. A 4-geon model of a
particle, originally constructed to explain quantum effects, is shown
to exhibit spin-half transformation properties. The well-known fact
that fermions transform like tethered objects is therefore explained
with an established classical theory - general relativity. It is
particularly significant that the explanation arises naturally, indeed
almost inevitably, as a consequence of the proposed gravitational
explanation of quantum mechanics \cite{hadley97,hadley98}.

\ack I am grateful to Dr M Micallef (Mathematics Institute,
University of Warwick) for valuable discussions and many helpful
suggestions.

\end{document}